\documentclass[10pt,conference,final]{./IEEEtran}

\IEEEoverridecommandlockouts
\usepackage{cite}
\usepackage{amsmath,amssymb,amsfonts}
\usepackage{algorithmic}
\usepackage{graphicx}
\usepackage{textcomp}
\usepackage{xcolor}
\def\BibTeX{{\rm B\kern-.05em{\sc i\kern-.025em b}\kern-.08em
		T\kern-.1667em\lower.7ex\hbox{E}\kern-.125emX}}

\usepackage{epsfig,endnotes}

\usepackage{booktabs} % For formal tables
\usepackage{subcaption}
\usepackage{multirow}
\usepackage{balance}

\begin{document}
	
\title{Efficient Attack Correlation and Identification of Attack Scenarios based on Network-Motifs}

%\author{
%	\IEEEauthorblockN{Anonymous Author(s)}
%	\IEEEauthorblockA{\vspace*{12pt}}
%}

\author{
	\IEEEauthorblockN{1\textsuperscript{st} Steffen Haas}
	\IEEEauthorblockA{\textit{University Hamburg}}
	\and
	\IEEEauthorblockN{2\textsuperscript{nd} Florian Wilkens}
	\IEEEauthorblockA{\textit{University Hamburg}}
	\and	
	\IEEEauthorblockN{3\textsuperscript{rd} Mathias Fischer}
	\IEEEauthorblockA{\textit{University Hamburg}}
}

\IEEEpubid{978-1-7281-1025-7/19\$31.00 ̃\copyright ̃2019 IEEE}

\maketitle

\begin{abstract}
% Introduction: Problem and Requirement
An Intrusion Detection System (IDS) to secure computer networks reports indicators for an attack as alerts. However, every attack can result in a multitude of IDS alerts that need to be correlated to see the full picture of the attack.
In this paper, we present a correlation approach that transforms clusters of alerts into a graph structure on which we compute signatures of network motifs to characterize these clusters.
%To correlate these attacks, we present an approach based on so-called network motifs that characterizes them.
A motif representation of attack characteristics is magnitudes smaller than the original alert data, but still allows to efficiently compare and correlate attacks with each other and with reference signatures. This allows not only to identify known attack scenarios, e.g., DDoS, scan, and worm attacks, but also to derive new reference signatures for unknown scenarios.
% Results
Our results indicate a reliable identification of scenarios, even when attacks differ in size and at least slightly in their characteristics. Applied on real-world alert data, our approach can classify and assign attack scenarios of up to 96\% of all attacks and can represent their characteristics using 1\% of the size of the full alert data.
\end{abstract}


\section{Introduction}
\label{sec:introduction}

% - bla
In our interconnected society, attacks on IT systems can have significant impact.
% - IDS
A common security practice is to deploy a network-based Intrusion Detection Systems (IDS) and/or a host monitoring and host IDS. Such IDS will output alerts when they detect security incidents. Subsequently, alert correlation can be leveraged to find relations among all alerts and to cluster them to attacks, e.g., based on alert similarity~\cite{Zhou2010}. 
% - CIDS
In recent years, the increasing network traffic challenges network-based intrusion detection with the data volume that needs to be analyzed. Because of that, scalable Collaborative IDSs (CIDS)~\cite{Vasilomanolakis2015} emerged. They deploy multiple IDS sensors in the network and collaboratively analyze traffic on the level of alert detection~\cite{Cordero2016} or alert correlation~\cite{Yegneswaran2004,Cai2005,Haas2018}. For that, IDS sensors either communicate with a central alert correlation unit or directly with each other.

% - Computation and Communication Overhead
Despite the benefits of larger visibility and load distribution that come with CIDSs, they introduces new challenges to the alert correlation process as individual sensors now need to exchange high-volume data in the form of alerts.
A pairwise comparison and correlation of all attacks and their alerts via a CIDS induces a high computation and communication overhead. For example, in a centralized CIDS, every sensor will share every single alert with the central alert correlation unit. To avoid this, an efficient and compact abstraction for attacks is required that can be precomputed by every sensor, shared efficiently, and that still allows for an efficient analysis of attacks and their interconnections. 
% -- Lower communication overhead
To achieve this, the attack abstraction needs to be significantly smaller than the original alert sets. Correlating attacks on the basis of their smaller abstractions, consumes significant less resources and therefore allows for faster and more efficient correlation algorithms. Thus, a CIDS benefits from a small abstraction through less data that needs to be exchanged between individual sensors or in between sensors and a central alert correlation unit.
However, such a reduction in size should preserve characteristic information on the attack.  Hence, these abstractions of alert sets should represent some kind of fingerprints that can be compared with each other, e.g., for similarity.
\IEEEpubidadjcol
% - Generic approach required
Most algorithms correlate attacks with respect to common attackers, i.e., on the basis of common IP addresses~\cite{Zhou2009,Locasto2005a,Haas2018}. However, a more generic approach is required that subsumes specific correlation algorithms and can be applied to a decentralized CIDS setting.
% -- Correlation according to scenarios
For that, correlating according to the same attack scenario might be equally important. An attack scenario is a class of attacks that all show a characteristic communication pattern among the involved hosts. The communication structure of a DDoS attack is different from a worm spreading and both differ from the lateral movement of an attacker in a network.

% - Contribution
The main contribution of this paper is a novel attack correlation algorithm that uses abstractions of attacks based on network motifs to identify attack scenarios. For that, we transform all alerts belonging to an attack into a graph representation from which we derive motif signatures. Motifs are characteristic subgraphs and a motif signature summarizes the occurrence of different types of motifs in a graph. It thus can serve as a fingerprint for the respective attack and does neither include IP addresses nor ports. This allows not only for a more compact attack representation, it is also more privacy-preserving as attack patterns can be shared without revealing identities of involved systems and services. That can help to reduce the load in a CIDS, as IDS sensors exchange motif signatures first and only when signatures match, i.e., they experience similar attacks, they would exchange more detailed information, e.g., complete alert clusters.
% -- Results
Our results indicate that we can correlate attacks from the same attack scenario with high accuracy and that our approach can also operate in a completely unsupervised setting to detect unknown attack scenarios. We apply our approach on real-world data and can represent the motif characteristics of attacks at 1 \% of the size of the full data and can classify and assign attack scenarios to up to 96 \% of all attacks.

% Roadmap
The remainder of this paper is structured as follows. In Section~\ref{sec:related_work} we present related work. Section~\ref{sec:system_chapter} describes our approach that incorporates motifs into an attack correlation algorithm for the identification of known attack scenarios and for learning new attack scenarios. The performance is evaluated in Section~\ref{sec:evaluation} and Section~\ref{sec:conclusion} concludes our work.

\section{Related Work}
\label{sec:related_work}
We present related work with respect to three areas. First, we look at alert correlation algorithms that cluster alerts to attacks and then classify these attacks based on predefined scenarios. Second, we look at approaches to data sharing in Collaborative Intrusion Detection Systems (CIDS). Last, we look at how network motifs have been previously used in the context of network security.

\paragraph{Attack Scenarios}
There are many alert correlation algorithms that extract the predominant information from a set of alerts, where each alert consists of attributes such as source and destination IP and port. The intention is to find groups of alerts within the larger set of alerts that share a common pattern and thus are likely to belong to the same underlying attack. These patterns usually consists of multiple pairs of attributes and their corresponding values which are equal for all alerts in the same cluster.
% Lattice - Pattern index
Zhou et al.~\cite{Zhou2009} describe how to reduce a large alert set to attribute patterns by using a lattice structure. Furthermore, they identified eight special types of patterns and assign them a so-called attack type. For example, a pattern with fixed source IP and port and fixed destination port is a distributed reflector DoS.
% GAC - Scenario matching with uncertainty
Instead of finding attribute patterns among the alerts, Haas et al.~\cite{Haas2018} propose community clustering on an alert graph. The authors then assign each resulting attack cluster one of four scenarios that are differentiated by the number of attackers and victims respectively.
% Zhu - ?
Zhu et al.~\cite{Zhu2006} automatically construct attack scenarios from alert data by modeling causal relationship of two alerts in neural networks.
% TCP State
Jero et al.~\cite{Jero2018} present a more specific approach to scenario detection. They leverage a state machine model of TCP congestion control to automatically generate abstract attack strategies.

\paragraph{Data Sharing in CIDS}
In CIDS, sensors are distributed in the network~\cite{Vasilomanolakis2015}. They perform intrusion detection locally~\cite{Mirsky2018} but exchange (parts of) their detection results. A low volume data exchange is usually desired for efficiency.
% Worm Containment
Cai et al.~\cite{Cai2005} propose a DHT-based overlay that is used by the sensors for detecting worms. For that, sensors generate signatures about suspicious packets locally and share them amongst each other. A worm is detected if the signature is observed by a sufficient amount of sensors in the CIDS.
% DOMINO
Domino~\cite{Yegneswaran2004} is an overlay system with multiple layers that was designed for the monitoring of Internet outbreaks. The so-called axis nodes are organized in a DHT-based overlay and form the backbone of the system. Satellite nodes form communities and report their alert data to axis nodes to share their aggregation in the axis overlay. These summaries describe attack data and are based on ports, sources, or alert clusters.
% AoI - Attribute patterns
Julisch~\cite{Julisch2003} presents an alert correlation approach based on attribute-oriented induction that aggregates attributes in an alert set to report attribute patterns similar to Zhou et al.~\cite{Zhou2009}. The resulting pattern with their corresponding values is an compression of the alert set itself.
% IP addresses in Bloom filters
Locasto et al.~\cite{Locasto2005a} propose a privacy-preserving compression of alert data for sharing suspicious IP addresses. They are inserted into a bloom filter, a bitmap that basically allows insertion and lookup operations of hashed input. This prevents anyone from retrieving the raw IP addresses but allows for lookups of a specific IP address. A compression in size is also achieved because multiple IPs can be inserted into the bloom filter that is of fixed size.

\paragraph{Network Motifs for Network Security}
% What are motifs?
A directed graph $G = (V,E)$ with its vertices $V$ and edges $E \subseteq V \times V$ can be characterized by network motifs~\cite{Milo2002}. They express how the vertices $V$ are connected among each other in $G$. For that, we look at individual sub-graphs $G' = (V', E')' \subseteq G$ that consist of a fixed number $n$ of vertices, i.e., $|V'| = n$ nodes from $V$ and out of edges $E' \subseteq E$.
\begin{figure}[h]
	\centering
	\includegraphics[trim={160 60 210 180},clip,width=0.75\linewidth]{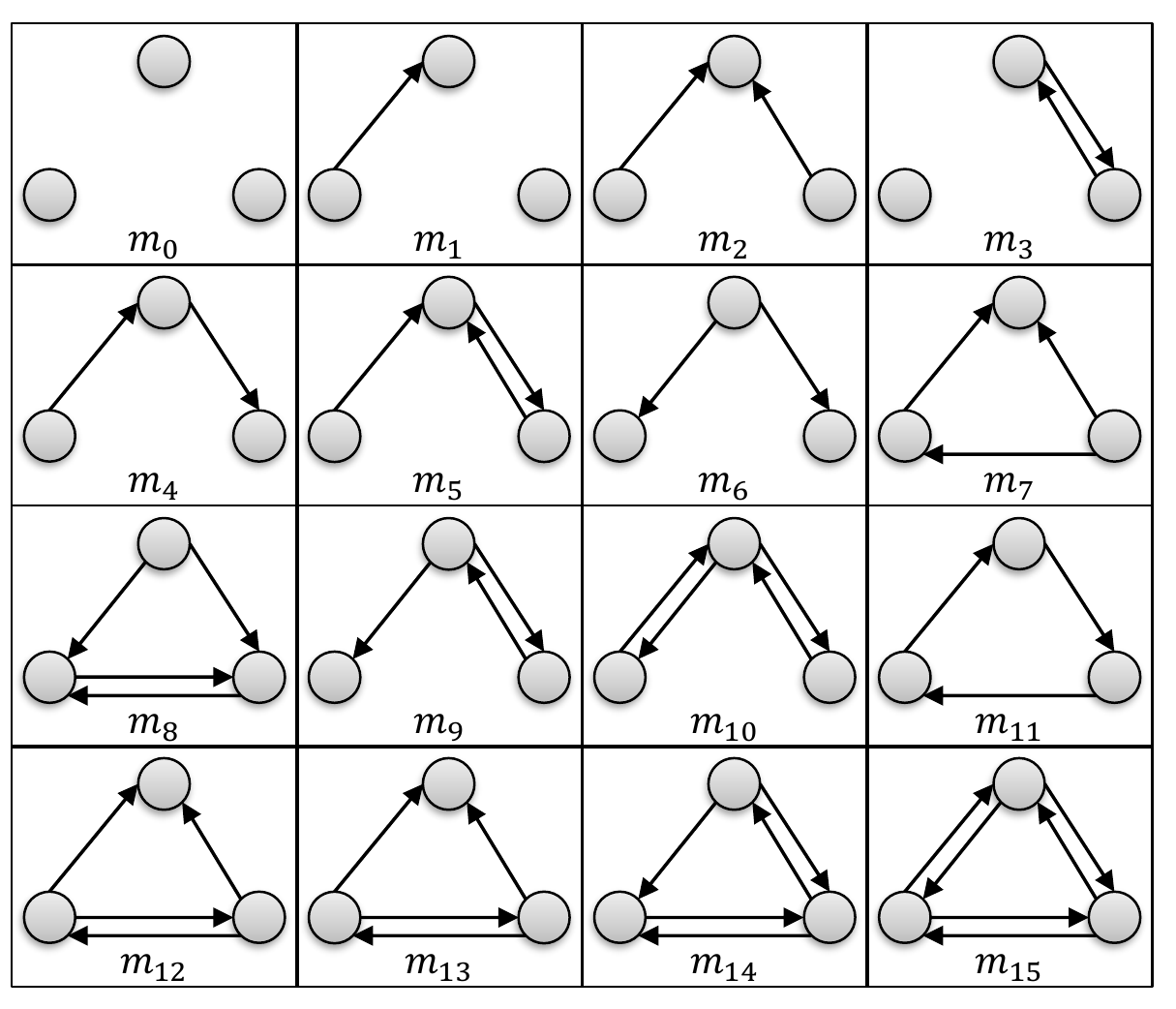}
	\caption{All 16 possible patterns for subgraphs of 3-motifs
	\label{fig:motif_instances}}
\end{figure}
% Motifs identify a subgraph pattern
The idea of motifs is based on the finite number of edges that can at most exist among any $n$ nodes, which is limited by $n \cdot (n-1)$ in case of a directed graph. This further means that $2^{n \cdot (n-1)}$ different graphs can be generated with $n$ nodes. Motifs in fact describe isomorphic classes for these possible graphs. Thus, a motif expresses a specific pattern how $n$ nodes could be interconnected. For example, for subgraphs of size $n=3$, $N=16$ distinct motif patterns exists, which are identified by their motif index $m_i$ with $i \in [0,N-1]$. Figure~\ref{fig:motif_instances} shows all possible motifs for $n=3$, where $m_0$ represents $n$ nodes without any edges and all $n \cdot (n-1)$ directed edges exist for $m_{N-1}$.
% Motif signatures
Thus, we can map any subgraph $G'$ to a motif index $m_i$. When we do this for all subgraphs $G' \subseteq G$ in our graph $G$ and count the occurences of every $m_i$, the result is called a motif-signature. Such a signature describes how often a specific motive is present in graph $G$.
% Identification of Application Protocols
With respect to network security, Allan et al.~\cite{Allan2009} are identifying application types in network traffic. With the help of motifs, they analyze the communication graph to predict the application.
% Visible changes in Motif Signatures for netflow graph during attacks
Juszczyszyn et al.~\cite{Juszczyszyn2011} show that motifs can be used to detect statistical anomalies in the communication graph during an attack in the network.
Similarly, Harshaw et al.~\cite{Harshaw2016} characterize the role of individual systems in the communication graph to detect the presence of an attack.

However, to the best of our knowledge, there is no related work that uses motifs in intrusion detection to identify attack scenarios. 
GAC \cite{Haas2018} can only identify four pre-defined cluster classes (oto, otm, mto, mtm). Our Motif-approach can characterize attacks generically, predefined, dynamically, and way more fine-grained. 
Compared to \cite{Juszczyszyn2011, Harshaw2016}, our data basis is on alert data and we use a novel graph model for representing communication relationships (preserving different usage behaviour of ports). The authors of \cite{Juszczyszyn2011, Harshaw2016} have shown that attacks change the NetFlow graphs significantly. In contrast, we (1) design a system that allows to compare attack characteristics, (2) evaluate the ability to differentiate between different attacks, and (3) create attack signatures that can be shared to identify similar attacks.

\section{Characteristics of Attack Scenarios}
\label{sec:system_chapter}
% Problem Description
With the increasing amount of attacks on computer networks and IT systems, there is a need to efficiently categorize, filter and correlate alerts to understand the root cause of an attack and to be able to choose appropriate countermeasures for its mitigation.
% Approach - Highlights
We present an approach based on network motifs~\cite{Milo2002} that provides abstractions of attacks, i.e., of their alert data. Our motif abstraction is a fixed size characteristic fingerprint of an attack. Thus, it can be magnitudes smaller than the corresponding alert data and allows for a faster comparison of attacks. With the help of this motif abstraction, our approach identifies attack scenarios and is even able to learn previously unknown scenarios. It can be deployed in a centralized or decentralized manner.

% Goal: Attack Scenarios
The goal is to inspect attacks in a way that differs from most algorithms that search for common attributes among the alerts from different attacks. These approaches can reveal if two victims are targeted by the same attacker but they cannot tell if the attacker performs the same kind of attack in both cases. That is why our approach identifies and compares attacks regarding their attack scenario, which basically describes how attackers and victims interact. If an attack is not only involving a single attacker and a single victim, the questions arises who of the involved hosts attacks whom. The answer can be simple for a Distributed Denial of Service (DDoS), where all attackers target one victim. But how to differentiate between a worm spreading and a coordinated scan, where each of the attackers scans a subset of victims?

% Outline
In the next Section~\ref{subsec:comparing_characteristics}, we describe the principles behind the classification of attack scenarios by calculating the similarities between motif signatures on the alert data of attacks. Afterwards in Section~\ref{subsec:attack_classification}, we describe how to utilize the comparison of motif signatures to classify known and unknown attack scenarios.

\subsection{Comparing Characteristics of Attacks}
\label{subsec:comparing_characteristics}

% Motifs for Attack Characteristics
The basic idea of our correlation algorithm is to transform a set of alerts into a much smaller representation that conserves structural characteristics of attacks. We found network motifs, specifically the so-called motif signatures (cf. Section~\ref{sec:related_work}), to be a perfect candidate to summarize the communication structure of the hosts involved in an attack. This allows for an easy comparison of the structural characteristics of attacks, even without prior knowledge of these characteristics and mostly independent from the attack size.

\begin{figure}[h!]
	\centering
	\includegraphics[trim={50 170 460 280},clip,width=0.8\linewidth]{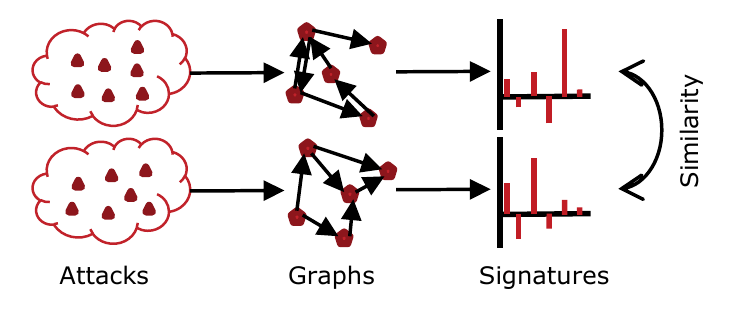}
	\caption{Schema to compare the alerts of two attacks by their motif signatures}
	\label{fig:motif_sim_overview}
\end{figure}

Please note that we present a correlation on attack level, which requires the alert data to be clustered in the preferred way of choice, e.g.~\cite{Haas2018,Julisch2003}, to extract clusters of alerts. This is also the first step according to the schematic overview of the approach in Figure~\ref{fig:motif_sim_overview}. In the following we describe the next steps in detail, which is the transformation of attack data into graphs, the calculation of motifs signatures, and their comparison.

\subsubsection{Transformation to Graph}
\label{subsubsec:graph_transformation}
% Communication Structure Graph
We assume attack characteristics to be based on the communication structure among hosts. This structure is derived from all alerts $a_i \in A$, where an alert $a_i$ has several attributes. For our purpose, we define an alert $a_i = (S\textnormal{:}T \rightarrow D\textnormal{:}L)$ with source IP $S$ and source port $T$ and with destination IP $D$ and destination port $L$. Based on these four attributes of alerts, we generate a \textit{Communication Structure Graph} $G_{com}$ for all alerts of an attack $A$. In $G_{com} = (V,E)$, nodes $v \in V$ represent either a host by its IP address or the port on a specific host. The edges reflect who attacked whom and whether a single port is relevant in the attack. 

\begin{figure}[h]
	\centering
	\includegraphics[trim={50 260 460 220},clip,width=0.7\linewidth]{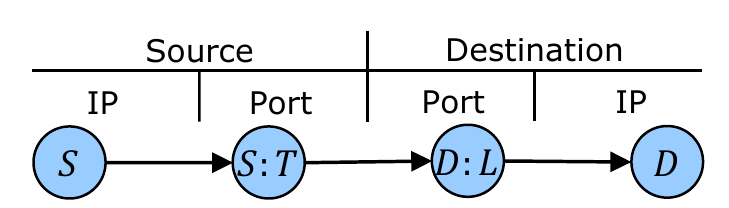}
	\caption{Adding an alert $(S\textnormal{:}T \rightarrow D\textnormal{:}L)$ to graph $G_{com}$}
	\label{fig:alert_to_graph}
\end{figure}

% How to build the graph
To build the graph $G_{com}$ for a specific attack $A$, all alerts $a_i \in A$ are added to it consecutively. For that, the set of nodes $V$ is extended by nodes representing the hosts, i.e., $\{S, T\}$, and nodes representing their ports, i.e., $\{S\textnormal{:}T, D\textnormal{:}L\}$. This notation ensures that ports are always bound to hosts. To reflect who attacked whom, the edges $\{(S, S\textnormal{:}T), (S\textnormal{:}T,D\textnormal{:}L), (D\textnormal{:}L,D)\}$ are added to $E$ in $G_{com}$. Intuitively, this describes what is visualized in Figure~\ref{fig:alert_to_graph}, the port and IP used to attack another IP on a respective port.

\subsubsection{Calculation of Signature}
\label{subsubsec:motif_signatures}
% From absolute count...
The last step to get from an alert set $A$ via the graph $G_{com}$ to its small abstraction is to calculate the motif signature of the graph (cf. Section~\ref{sec:related_work}). For that, we enumerate all subgraphs $G' = (V',E') \subseteq G_{com}$ of size $|V'| = n$ and assign them to a specific motif pattern $m_i$ with $i \in [0,N-1]$.
Counting the number of occurrences for all motifs $m_i$ results in a motif signature $M^A$, which is a vector containing the absolute number of occurrences of every $m_i$. As $M^A$ is directly dependent on the graph size $|G_{com}|$, it does not allow to compare the structure of two graphs that are of different sizes.

% ... to Z-Score
To enable this comparison, the authors in~\cite{Milo2002} present the so-called Z-Score. It uses the signature $M^A$ of graph $G_{com}$ to calculate for every $m_i$ how much it is over- or underrepresented compared to a random graph of the same size and with the same number of edges as $G_{com}$. The Z-Score of a specific motif $m_i$ in a graph $G$ is calculated by $Z(m_i) = \frac{M^A(i) - M^{rand}(i)}{sd}$ with $M^A(i)$ being the absolute number of motif $m_i$ in $G_{com}$, $M^{rand}(i)$ being the average absolute count of motif $m_i$ in respective random graphs, and its standard deviation $sd$. This is done for every absolute number of motifs in $M^A$. We denote the resulting motif signature with the Z-Score values as $M^Z$. Any motif signature can simply be represented as an array of fixed length, e.g., 16 for 3-motifs.

\subsubsection{Comparison of Signatures}
\label{subsubsec:signature_comparison}

% Values in Signatures
When comparing two motif signatures $M^Z_1$ and $M^Z_2$, we want to calculate how similar they are. Their similarity should be 1, i.e., $100\%$, if the signatures are equal. However, finding a metric to calculate the similarity in a meaningful manner is not intuitive. The reason is because even the values in the Z-Score signature $M^Z$ are not limited to a fixed range. As the graph size, i.e. attack size, still has an impact on the Z-Score values, we cannot directly compare the values of every motif in two signatures $M^Z_1$ and $M^Z_2$. Doing so would not achieve high similarity for attacks with similar characteristics but of different size.
% Shape of Signature
We have to design the comparison between two motif signatures in such a way that similar attack characteristics are identified even if the attacks are of different sizes. For that, a motif that is statistically over- or underrepresented in $M^Z_1$ should also be statistically over- or underrepresented in $M^Z_2$. Furthermore, we make the comparison to consider how much a specific motif is over- or underrepresented compared to the other motifs in the signature. The idea is to not have a pairwise comparison of the motif values in $M^Z_1$ and $M^Z_2$. Instead, the similarity between $M^Z_1$ and $M^Z_2$ reflects how similar the relations among the motif values in $M^Z_1$ are to the relations among the motif values in $M^Z_2$.

% Vector in Space
For such a comparison, the Z-Score signatures $M^Z_1$ and $M^Z_2$ are interpreted as vectors $\vec{u}, \vec{v}$, always of fixed length. This means, we can think of a signature as a vector in a multi-dimensional space with the number of dimensions equal to the length of the vectors. The higher the values in a signature, the larger is the vector in the multi-dimensional space. 
% Angle between two vectors
To be independent from the vector length, we calculate the angle between two vectors in the multi-dimensional space. For that, we use the inner product $<\vec{u} ,\vec{v}>$ and the Euclidean norms $||\vec{u}||_2$ and $||\vec{v}||_2$ for the calculation of the angle $\phi$ (Equation~\ref{equ:vector_degree}). This finally leads to the similarity $0 \leq sim \leq 1$ (Equation~\ref{equ:vector_sim}).
\begin{equation}
	cos(\phi) = \frac{<\vec{u},\vec{v}>}{||\vec{u}||_2 \cdot ||\vec{v}||_2}
\label{equ:vector_degree}
\end{equation}
\begin{equation}
	sim = \frac{cos^{-1}(\phi)}{\pi}
\label{equ:vector_sim}
\end{equation}

% Minimum Similarity
When we compare the Z-Score signatures $M^Z$ to classify attack scenarios, we define a threshold $\tau$ for their minimum similarity. Above this threshold, two signatures belong to the same attack scenario. Next, Section~\ref{subsec:attack_classification} makes use of this threshold when testing if an attack matches a given attack scenario or when clustering attacks for unknown attack scenarios.

\subsection{Classification of Attacks}
\label{subsec:attack_classification}

% Comparing Motif signatures
Once the alerts of every attack are transformed into motif signatures $M^Z$, the motif signatures of the different attacks, i.e., alert sets, are compared. The goal is to find attacks with the same characteristics. Such a comparison can be calculated extremely fast as a motif signature is of a fixed and far smaller size than the respective alert sets.
% - Identifying Attack Scenarios
For the attack correlation algorithm, predefined characteristics of attacks can be used to identify specific attack scenarios and to label them accordingly. 
% - Learning Attack Scenarios
Alternatively, the attack correlation operates without a knowledge database and learns attack scenarios on its own. It labels attacks according to dynamically derived characteristics of alert sets. Both approaches are described in the following.

\subsubsection{Signature-based Classification}
\label{subsubsec:signature_classfication}
% Reference Scenarios
To identify scenarios in alert sets, we first need definitions for the characteristics of already known attack scenarios that we name reference scenarios $R_i \in \mathcal{R}$. Thus, a reference scenarios $R_i$ reflects the characteristics of a specific attack scenario and therefore is representative for all attacks of this attack scenario and their alert sets, respectively. In fact, $R_i$ is just the motif signature $M^Z$ of a typical attack in the attack scenario that should be identified. It is modeled by transforming a representative alert set $A$ into a \textit{Communication Structure Graph} $G_{com}$ and by computing its motif signature $M^Z$. Hence, set $\mathcal{R}$ consists of small motif signatures. The size of this set equals the number of predefined attack scenarios that should be identified.

% Highest Similarity Wins
When classifying an attack $A$, its motif signature $M^Z$ is compared to all reference signatures $R_i \in \mathcal{R}$ using the similarity function described in Section~\ref{subsubsec:signature_comparison}. In general, the highest similarity determines the attack scenario that is assigned to attack $A$. However, the minimum similarity threshold $\tau$ needs to be respected, because there could be attacks from unknown scenarios that are not included in $\mathcal{R}$. Hence, they should not be labeled with a known attack scenario. When assigning all attacks to the reference scenarios, it results in attack clusters $C_x \in \mathcal{C}$, at most one per reference scenario. The requirement for attacks of the same scenario $M^Z_i \in C_x$ to have a minimum similarity $\tau$ to the respective reference scenario $R_x$ is formalized in Equation~\ref{equ:reference_min_sim}. The requirement of closest match among all reference scenarios $\mathcal{R}$ is formalized in Equation~\ref{equ:reference_closest_sim}.
\begin{equation}
	\forall M_i \in C_x: sim(M_i, R_x) \geq \tau
\label{equ:reference_min_sim}
\end{equation}
\begin{equation}
	\forall M_i \in C_x: \forall R_j \in \mathcal{R}: sim(M_i, R_x) \geq sim(M_i, R_j)
\label{equ:reference_closest_sim}
\end{equation}

\subsubsection{Unsupervised Clustering}
\label{subsubsec:dynamic_classfication}
% Learning attack scenarios
Apart from using motif signatures to identify predefined attack scenarios, we explain how motifs can be used to cluster similar attacks and to dynamically derive reference scenarios $\mathcal{R}$ for them. 
In the following, we describe the process of learning new attack scenarios via a hierarchical clustering in two steps. This only highlights the capabilities of motifs for applications in intrusion detection. The fundamentals presented here should be easily adaptable to more sophisticated detection algorithms, e.g.\ for anomaly detection or collaborative intrusion detection.

\paragraph{Hierarchical Clustering}
% Clustering on Similarity
To learn attack scenarios, we first cluster attacks in a way such that all attacks from the same attack scenario belong to one group. For that, we cluster attacks based on the similarity of their motif signatures $M^Z$. We designed the comparison of motif signatures $M^Z$ to aim for high similarities among signatures for attacks of the same scenario and low similarities of signatures for attacks from different scenarios. The intention is that clustering the motif signatures of attacks results in attack clusters, one for each detected attack scenario.
% Hierarchical clustering
We find hierarchical clustering best to cluster attacks into attack scenarios for two reasons. First, we can use the similarity threshold $\tau$ (c.f. Section~\ref{subsubsec:signature_comparison}) as clustering parameter to intuitively control the clustering and its outcome. And second, the visualization as dendrogram allows an human inspection of the potential clusters depending on $\tau$.

\begin{figure}[h]
\centering
\includegraphics[trim={5 20 10 10},clip,width=0.9\linewidth]{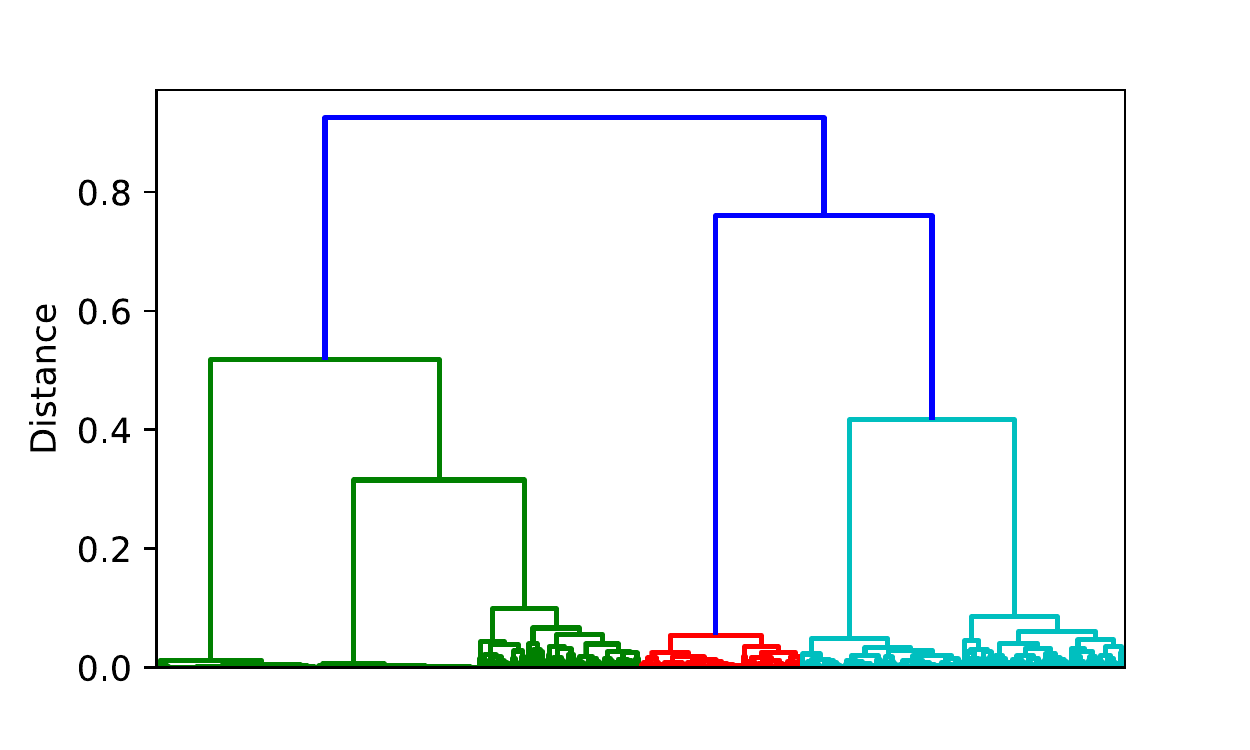}
\caption{Hierarchical clustering for attacks from six example scenarios}
\label{fig:dendrogram_popFact10}
\end{figure}

% Similarity Parameter
We use $\tau$ as clustering parameter in hierarchical clustering to control the minimum similarity, i.e., maximum distance, for two attacks, so that they are still part of the same attack scenario. Thus, $\tau$ must be chosen in a way such that (1) it is low enough to allow attacks of the same scenario to result in one cluster and (2) it is high enough that two attacks of different scenarios do not result in the same cluster.
% Dendrogram
An example for hierarchical clustering of attacks is visualized as dendrogram in Figure~\ref{fig:dendrogram_popFact10}. The x-axis of the figure represents the individual attacks and the y-axis represents $1 - \tau$ as maximum distance for two attacks or scenarios to be merged into the same scenario. Hence, at distance $0$, only equal attacks will be merged and at distance $1$, all attacks are merged into a single scenario. When stepping from $0$ to $1$, similar attacks or scenarios are merged once the value on the y-axis reaches their respective distance.
The idea in hierarchical clustering is to define a cut-off distance, which determines the final clusters. The outcome are the clusters, i.e.\ scenarios, that all the attacks have been merged into at the specific cut-off distance in the dendrogram.

We use the cut-off distance $1-\tau$ to form clusters $C_x \in \mathcal{C}$ of attacks $M^Z_i \in C_x$ with the desired maximum heterogeneity within a cluster. The hierarchical clustering of $M^Z$ for the set of attacks merges clusters with respect to the maximum distance within the resulting cluster, which is known as the \texttt{complete} method~\cite{Sorensen1948}. This means that attack scenarios are defined by the maximum distance between contained attacks, which can be formally stated as in Equation~\ref{equ:clustering_homogeneity}.
\begin{equation}
\label{equ:clustering_homogeneity}
	\forall C_x \in \mathcal{C}: \forall M_1, M_2 \in C_x: sim(M_1, M_2) \ge \tau
\end{equation}

\paragraph{Deriving Reference Scenarios}
% Specific attack as referece scenario
The attacks in a cluster $C_x \in \mathcal{C}$ formed by hierarchical clustering are supposed to belong to the same attack scenario because of their similar characteristics. To actually extract the characteristics for each attack cluster, a motif signature $M^Z$ is derived per cluster as reference scenario $R_i \in \mathcal{R}$ to represent the new attack scenario. Instead of constructing a signature ourselves, we pick the signature of the attack that is most typical for the cluster. We define the most typical attack of a cluster to have the highest similarity with every other attack on average. This is formalized in Equation~\ref{equ:derived_signature}.
\begin{equation}
\label{equ:derived_signature}
	\forall M_i \in C_x: \sum_{j=0}^{|\mathcal{C}_x|}sim(M_i, M_j) \leq \sum_{j=0}^{|\mathcal{C}_x|}sim(R_i, M_j)
\end{equation}

After these two steps, a set of attacks, i.e., a set of alert sets, is represented by a number of reference scenarios that is controlled by the similarity threshold $\tau$.

%\shci{Summary}

\section{Evaluation}
\label{sec:evaluation}

% General Implementation
For evaluating our correlation algorithm (cf. Section~\ref{sec:system_chapter}), we implemented it in Python. The implementation encompasses the graph generation, motif calculation and comparison as well as the identification of known attack scenarios (cf. Section~\ref{subsubsec:signature_classfication}) and the detection of unknown scenarios (cf. Section~\ref{subsubsec:dynamic_classfication}).
Prior to the description of the experiments and their results, we introduce the alert datasets used for the experiments in Section~\ref{subsec:attack_data}.

% Experiments
The experiments are divided into two parts. First, Section~\ref{subsec:signature_evaluation} evaluates the applicability of motif signatures to identify and compare different attack scenarios. By simulating attacks, we show that our correlation algorithm can correctly classify attack scenarios, mostly independent from the attack sizes.
The second part in Section~\ref{subsec:real_world_evaluation} then evaluates our correlation algorithm on real-world data.

\subsection{Attack Data}
\label{subsec:attack_data}

Although our correlation algorithm operates on alerts, it is important to notice that the algorithm works with abstractions of attacks. Hence, the input to our correlation algorithm are alerts of specific attacks, i.e., clusters of alerts that each represents an individual attack. For clustering alerts to attacks, we refer to other algorithms, e.g.,~\cite{Valdes2001,Haas2018,Julisch2003,Zhu2006,Pei2016}.
In the experiments we use two different types of alert data:
\begin{itemize}
	\item \textbf{Synthetic} alerts of attacks with ground-truth and control over to carefully evaluating the classification accuracy of our approach.
	%(Section~\ref{subsubsec:attack_scenarios})
	We further use this method of alert generation for the construction of reference scenarios (cf. Section~\ref{subsubsec:signature_classfication}).
	\item \textbf{Real-world} alerts from DShield\footnote{https://secure.dshield.org}.
	%as described in Section~\ref{subsubsec:dshield_data}.
\end{itemize}
Next, we explain how to generate instances of attack scenarios with the help of synthetic alerts. Afterwards, we introduce the DShield data set as a source for real-world alerts.

\subsubsection{Instances of Attack Scenarios}
\label{subsubsec:attack_scenarios}

To create synthetic attack data, we generate alerts for a specific attack scenario. In our experiments we use six different scenarios, each of them defining a pattern for the data generation. An instance of an attack is determined by its attack pattern and its attack size. The values for IP and ports are randomly chosen from the full IP and port range, respectively.

% Creation patterns
\paragraph{Attack Patterns}
For our experiments, we describe the alert patterns for six attack scenarios. We use these patterns for two purposes: 1) For generating synthetic alerts of attacks for the evaluation of scenario classification and 2) for the definition of reference scenarios used during classification. The names and characteristics of the six attack scenarios are as follows:

% DDoS
A Distributed Denial of Service (\textbf{DDoS}) attack is characterized by alerts that all share the same destination IP and port. Thus, multiple attackers target a specific host and service. We can parameterize this attack by $\alpha$ alerts that are generated on average per attacker. Attackers use random source ports, which are reused in subsequent alerts with a probability $p$. 
% Scan
A \textbf{Scan} attack is characterized by alerts that share the same attacker IP. Random source ports are used to scan for the same destination port on multiple target machines. On average $\alpha$ alerts are generated per target and the attacker reuse a source port with a probability $p$.
% D-Scan
A Distributed Scan (\textbf{D-Scan}) is similar to a Scan attack but with both multiple attackers and targets, e.g., when a Scan is coordinated by a botnet~\cite{Haas2016}. Then, tasks for scanning all targets are split among the attackers. Characteristic for this scenario is the ratio of attackers to targets, which we denote as $\theta$. Additionally, targets might be scanned multiple times, i.e., from multiple attackers. In this case, $\alpha$ alerts are generated per target.
% Worm
A \textbf{Worm} attack is characterized by alerts that all share the same destination port. Additionally, all hosts are attackers and target randomly $\mu$ of other hosts via random source ports.
% Exploration
An Exploration (\textbf{Expl}) attack is characterized by a single attacker that targets $f$ hosts. Each compromised host serves as source for attacks on further hosts. Each compromised host targets $f$ new hosts and all source and target ports are random.
% Convergence
The pattern of an Convergence (\textbf{Conv}) attacker is the opposite of an Exploration attack. The actual target is attacked by $f$ hosts that themselves are attacked by $f$ hosts and so on.

% Parameters
For our experiments, we parameterize these six attack patterns as follows. Per source or target, $\alpha = 1.5$ alerts are generated. Ports are reused with a probability $p = 50\%$ where appropriate. In case of lateral movement in a network, it is done with a spread factor of $f = 5$, which means that in each step a new compromised host targets 5 new hosts. If a scenario is characterized by multiple attackers and targets, their ratio is $\theta = 0.5$, which means the same amount of attackers and targets. In the case of the worm scenario, each host attacks $\mu = 10\%$ of the other hosts. 
%Illustrations of the graph $G_{com}$ and the motif signature $M^Z$ for the six attack scenarios can be found in Appendix A.

\paragraph{Attack Variations}

In reality not all attacks of the same attack scenario are equal with respect to their alerts. Of course they differ in the actual IPs and ports, which, however, are not visible anymore in the motif signature of the attack. More interesting is the variation in the attack size. Also, clustered alerts of an attack can contain false positives which causes variations in the alerts of an attack. Attack variations of the six scenarios are input to the experiments in Section~\ref{subsec:signature_evaluation}.

% Size
We cause attack variations by generating attacks of different size, i.e., the \textbf{population} $\psi$, which is the number of hosts involved in the attack. The attack patterns define how many of the individual hosts are attackers, targets, or both. For example, in a DDoS attack of size 100, there would be one target and 99 attackers.

\subsubsection{DShield Data Set}
\label{subsubsec:dshield_data}

% DShield alerts
We are also using real-world data from the Internet Storm Center\footnote{SANS Technology Institute, Internet Storm Center, https://isc.sans.edu} that operates DShield, which is a platform for sharing data from security devices, e.g., from firewalls. The DShield logs consist of alerts from multiple sensors around the globe. For our real-world experiments, we are using all alerts collected on August, 22th in 2016. These are 4,517,497 alerts in total and are a result of several attacks.

% Clustering with GAC
As our approach for scenario classification works on attacks, i.e., on alerts of the same attack, we first have to group the DShield alerts into clusters. For that, we are using the GAC clustering approach~\cite{Haas2018} that clusters alerts based on attribute similarity. In contrast to other clustering approaches, e.g.,~\cite{Zhou2009,Julisch2003}, GAC does not enforce clusters with static attribute patterns. Instead, it identifies cliques of alerts that form a community, which allows a high diversity in the alert clusters and therefore in the attack scenarios. If we would use clustering algorithms with static attribute patterns, alert clustering would not be able to produce alert clusters for certain attack scenarios.
% Clustering results
Applying GAC clustering with a minimum similarity of $0.25$ in between alerts and a clique size of $15$ on the DShield data results in 34,204 clusters. They are the actual input to the real-world evaluation in Section~\ref{subsec:real_world_evaluation}.

\subsection{Classification of Attack Scenarios}
\label{subsec:signature_evaluation}

% Representation of Attack Scenarios
The most important question is if our motif-based classification can fulfill the requirements to our classification problem (cf. Section~\ref{sec:introduction}). We require the abstraction of attacks to be small and the fingerprinting of attacks to be characteristic for their scenarios. The fulfillment of the first requirement is given, because the data volume, i.e., size of all alert data can be magnitudes larger than the size of motif signatures. Thus, we investigate if a single motif signature is representative for all variants of an attack scenario.
% Intra-Class-Similarity
For that, motif signatures have to be very similar for attacks of the same attack scenario. We denote this as \textit{intra-class-similarity}, which describes the similarities among attacks from the same attack scenario.
% Inter-Class-Similarity
In addition, we also have to look at the \textit{inter-class-similarity}, which is the similarity between attacks from different attack scenarios. 
%The intra-class-similarity is required to be high for all attack scenarios. The inter-class-similarity is required to be low such that different attack scenarios can be distinguished.

\subsubsection{Similarities of Scenario Classes}
\label{subsubsec:scenario_similarities}

% Setup
In this experiment, we investigate the intra-class similarities and inter-class similarities for attacks of the same size $\psi = 100$, i.e.,  number of hosts. For that, we created 1000 attacks for each of the six scenarios defined in Section~\ref{subsubsec:attack_scenarios} and measured their similarities.
Please note that the attack patterns themselves inherent some randomness, so that two attacks of the same scenario differ in their alert data even if they are of equal size. Apart from different IPs and ports, the relation for who of the attackers target whom of the victims is chosen differently every time an attack instance is generated. More randomness is introduced, because of the generation parameter $\alpha = 1.5$ multiple alerts are generated for some attackers, which also differs every time.
% Similarities in numbers
\begin{table}[h]
	\centering
	\begin{tabular}{c|ccc}
		& \textbf{Lowest Intra-Class} & \multicolumn{2}{c}{\textbf{Highest Inter-Class}} \\
		& \textbf{Similarity [\%]} & \multicolumn{2}{c}{\textbf{Similarity [\%] (with)}}  \\\hline\hline
		
		\textbf{DDoS} & 99.64 & 78.77 & (Worm) \\
		\textbf{Scan} & 99.29 & 73.58 & (Worm) \\
		\textbf{D-Scan} & 88.26 & 73.42 & (Conv) \\
		\textbf{Worm} & 89.98 & 78.77 & (DDoS) \\
		\textbf{Expl} & 92.65 & 73.00 & (D-Scan) \\
		\textbf{Conv} & 91.52 & 73.42 & (D-Scan) \\
	\end{tabular}
	\caption{Similarities for attacks with 100 hosts for six scenarios, both for the same and for different scenarios.}
	\label{tab:class_similarities}
\end{table}

% Results
Table~\ref{tab:class_similarities} tells us how different attacks from an individual scenario might be, i.e., the lowest intra-class similarity. The highest variation is measured among the D-Scan attacks (lowest similarity of 88.26\%) and the most similar attacks from the same scenario are the DDoS and Scan attacks, each with a similarity of more than 99\%.
We also measured the inter-class-similarities and report the highest similarities per scenario in Table~\ref{tab:class_similarities}. There are some attack scenarios that share characteristics. The worm is similar to DDoS and Scan with 78.77\% and 73.58\%, respectively. The D-Scan is similar to Expl and Conv with 73.00\% and 73.42\%, respectively.

% Summary
As the lowest intra-class similarity is always higher than the highest inter-class similarity, motifs are an appropriate abstraction for alert data to preserve the characteristics of attack scenarios.
For attacks of the same size, the results indicate that our approach can correctly classify attack scenarios, both in sense of identifying the correct reference scenarios and detecting scenarios in unsupervised clustering.

\subsubsection{Scaling with Attack Size}
\label{subsubsec:exp_popFact}
% Intra- and Inter-Class-Similarities depending on number hosts
Apart from the question if the intra-class-similarity is always higher than the inter-class-similarity, the influence of the size of an attack is of interest. This is important because attacks can greatly differ in their sizes, i.e., number of hosts $\psi$. Our motif-based classification is required to detect the attack scenario for an attack with 100 hosts but also for an attack with 1000 hosts.
% Setup
For that, we generate attacks for the six attack scenarios described in Section~\ref{subsubsec:attack_scenarios} with different attack sizes. For the generated attacks, we calculate the intra-class similarity or the pairwise inter-class similarity, respectively, depending on if two attacks are from the same scenario or not.
% - Variation of size
We calculate the similarities for different sets of attacks. In the set of attacks, we control the difference between the size of smallest and largest attack, i.e., how different the attacks are with respect to their size. The smallest attacks always encompass 100 host and the largest attacks are of size up to 1000.

% Figures description
\begin{figure}[h]
	\centering
	\includegraphics[width=1.0\linewidth]{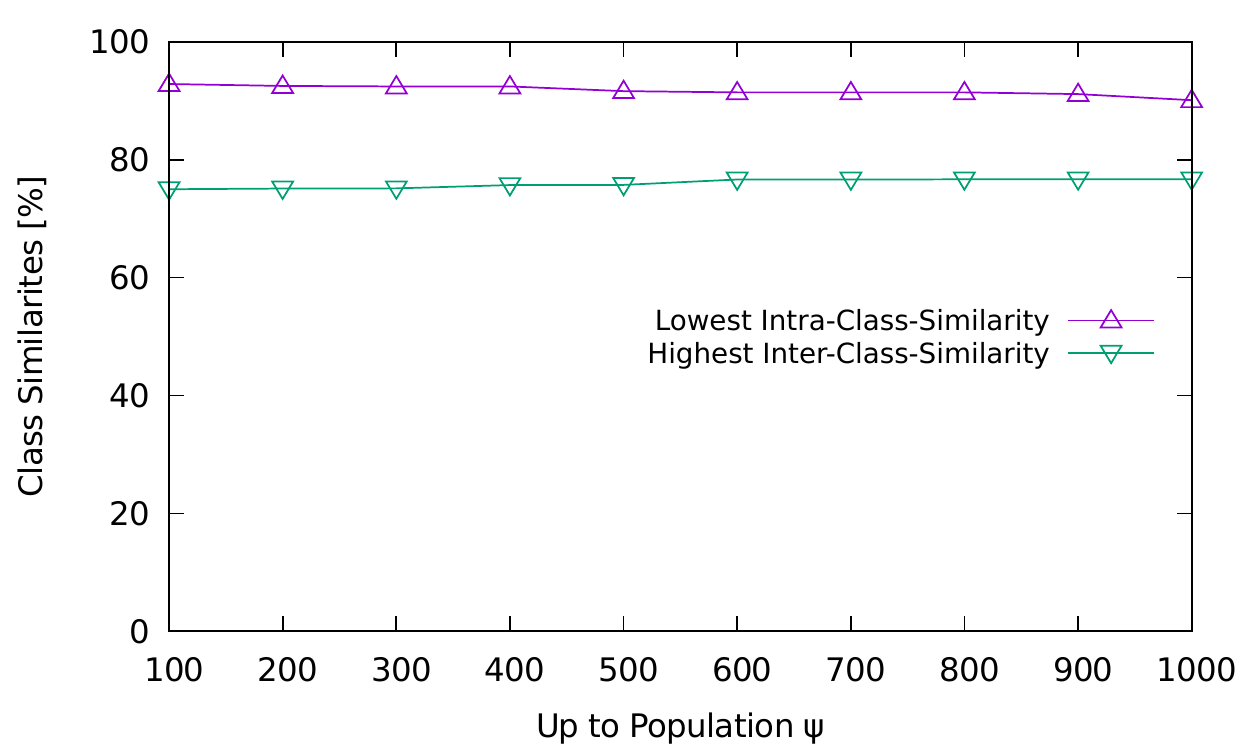}
	\caption{Distance between highest inter-class-similarity and lowest intra-class-similarity for attacks of several sizes. The range for attack sizes is $[100; \psi]$ and the similarities depend on the upper bound $\psi$.}
	\label{fig:sims_popFact_short_tolerances}
\end{figure}
\begin{figure*}[t]
	\begin{subfigure}[b]{0.49\textwidth}
		\centering
		\includegraphics[width=\textwidth]{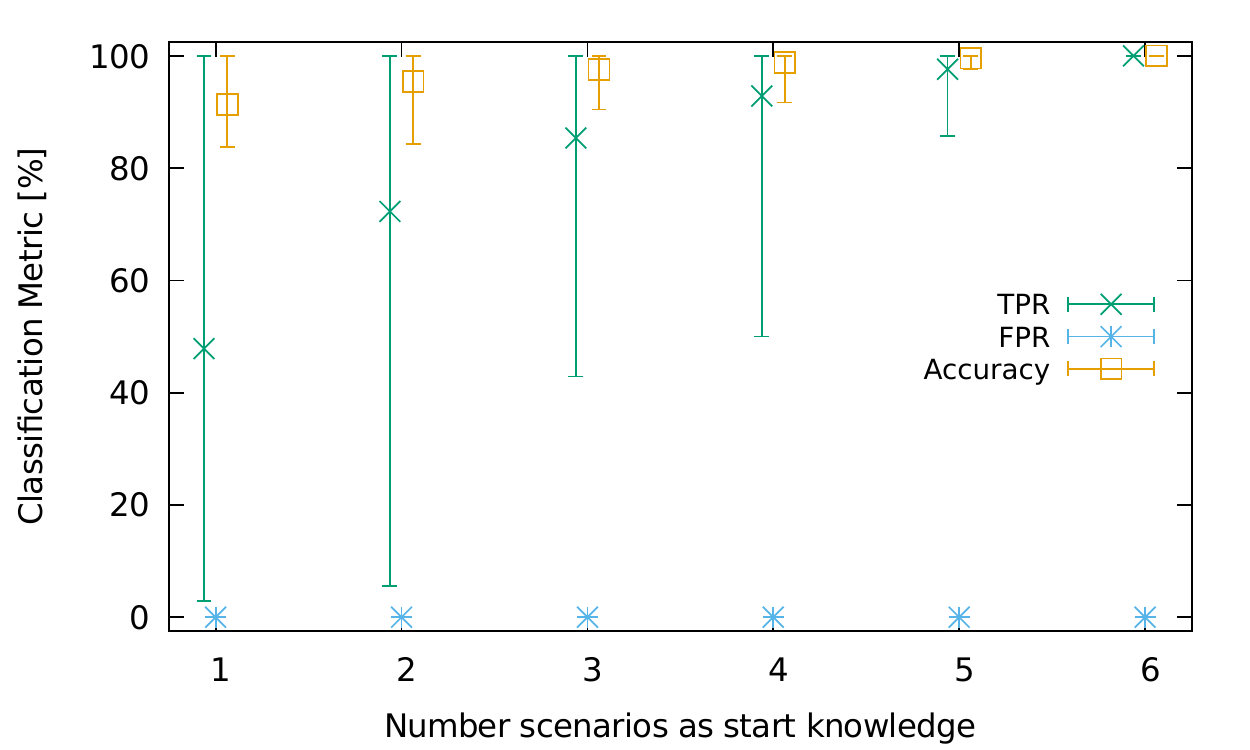}
		\caption{Based on lowest Intra-Class-Similarities}
		\label{fig:start_knowledge_intra_classes}
	\end{subfigure}
	\begin{subfigure}[b]{0.49\textwidth}
		\centering
		\includegraphics[width=\textwidth]{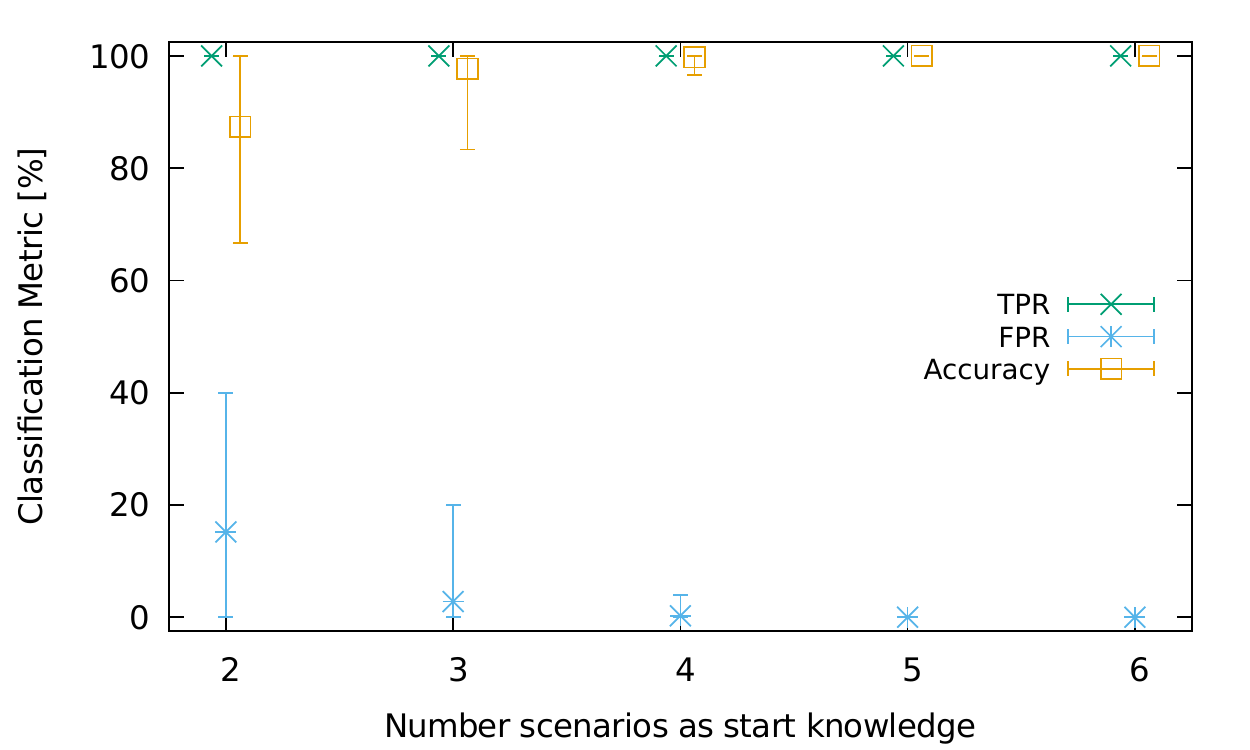}
		\caption{Based on highest Inter-Class-Similarities}
		\label{fig:start_knowledge_inter_classes}
	\end{subfigure}
	\caption{Evaluation of attack clustering and classification for attacks of sizes between 100 and 1000. The clustering similarity $\tau$ and the resulting classification metrics depend on the selection of known scenarios as start knowledge.}
	\label{fig:start_classificaton}
\end{figure*}
In Figure~\ref{fig:sims_popFact_short_tolerances}, we evaluate the similarities depending on the attack sizes, i.e., when increasing the range of attack sizes in steps of 100.
On the x-axis is the upper bound of the population size $\psi$, meaning a value on the y-axis depending on a specific $\psi$ plots the similarities among attacks of sizes $[100; \psi]$ in steps of 100. On the y-axis, we plot the lowest intra-class-similarity and highest inter-class-similarity among all attack scenarios. As long as the first curve is above the second one, it is possible to correctly classify the attacks in our data set.
% Threshold range of \tau
The gap between both curves indicates the potential range for the classification parameter $\tau$ to achieve correct classifications.
% Results
The results indicate that our motif-based approach can preserve the attack characteristics mostly independent from the attack size. For attacks of size 100 only, the width of the range for $\tau$ is $0.18$. When clustering data sets that contain attacks of sizes between 100 and 1000, the width slowly decreases to $0.13$.
With respect to attack sizes between 100 and 1000 and with respect to our six attack scenarios in this experiment, the average value of $\tau$ should be about $0.83 \textnormal{ +/- } 0.07$.

\subsubsection{Learning new Scenario Classes}
\label{subsubsec:exp_learning}

% Dynamic Scenario Classes depening on tau
Another question, especially regarding how to choose $\tau$, is how the accuracy of learning new scenarios depends on general knowledge of attack scenarios. While the unsupervised algorithm (cf. Section \ref{subsubsec:dynamic_classfication}) can detect and characterize new scenarios, it is a matter of operating the attack clustering with an appropriate value for $\tau$, not of specific previously defined reference signatures. For different choices of setting $\tau$, we show how deriving unknown scenarios performs.
% Scenario Classses
In this paper, we define only six attack scenarios but our approach is not limited to them. As our attack abstraction enables the identification and comparison of structural characteristics of attacks, our approach can potentially learn any new scenarios as long as they sufficiently differ in their communication structure. Another potential goal is to divide known scenarios into more fine-grained ones, i.e., differentiate a reflection DDoS from a DDoS performed by a botnet. However, for simplicity we only define the six general scenarios here and leave more (fine-grained) scenarios to future work. 
To still investigate the learning of new scenarios in this experiment, we assume a knowledge of only a subset of the six scenarios.

% Assumption of how tau influences TPR/FPR
According to the results illustrated in Figure \ref{fig:sims_popFact_short_tolerances}, the more $\tau$ is towards the upper bound of the possible range, the definition of attack scenarios becomes more strict. This results in a higher probability for false negatives in classifications in case the attacks of a new scenario have a higher variation than the previously known ones. However, a high $\tau$ also ensures precise classifications by reducing false positives in classifications in case the attacks of a new scenario share characteristics with a previously known one. If $\tau$ is set towards the lower bound of the range, there is a higher chance that attacks will be correctly classified although they look different than expected by the attack scenario. In turn, this increases the likelihood of false positives.

% Setup
For this experiment, we use the same data set as in Section \ref{subsubsec:exp_popFact}, containing attacks from the six scenarios with 100 to 1000 hosts. We then simulate limited knowledge by only considering the attacks from a subset of scenarios. For each combination of 1 to 6 scenarios, we measure the lowest intra-class and highest inter-class similarities as in Table \ref{tab:class_similarities} and then determine the highest and lowest possible $\tau$ for the attack scenarios as in Figure \ref{fig:sims_popFact_short_tolerances}.
As there are multiple possible combinations of scenarios per number of scenarios, Figure \ref{fig:start_classificaton} plots the minimum, maximum, and average value for the true-positive-rate (TPR), false-positive-rate (FPR), and accuracy of classification. As $\tau$ is based on the lowest known intra-class similarity in Figure \ref{fig:start_knowledge_intra_classes}, attacks from all remaining scenarios will definitively go to another class, i.e., we accept a lower TPR but therefore minimize the FPR for learning new scenarios. In contrast, Figure \ref{fig:start_knowledge_inter_classes} is based on the highest known inter-class similarity, which maximizes the TPR but accepts a higher FPR in return. Thus, choosing $\tau$ from the higher or lower boundary of possible range balances the ratio between expected TPR and FPR.

% Hybrid Learning
The results here for unsupervised clustering describe the worst-case performance of attack classification. For signature-based classification (cf. Section \ref{subsubsec:signature_classfication}), attacks are only required to be more similar to the reference signature of their respective scenario than to the reference signature of a different scenario.

\begin{figure*}[t]
	\centering
	\begin{subfigure}[b]{0.49\textwidth}
		\centering
		\includegraphics[width=\textwidth]{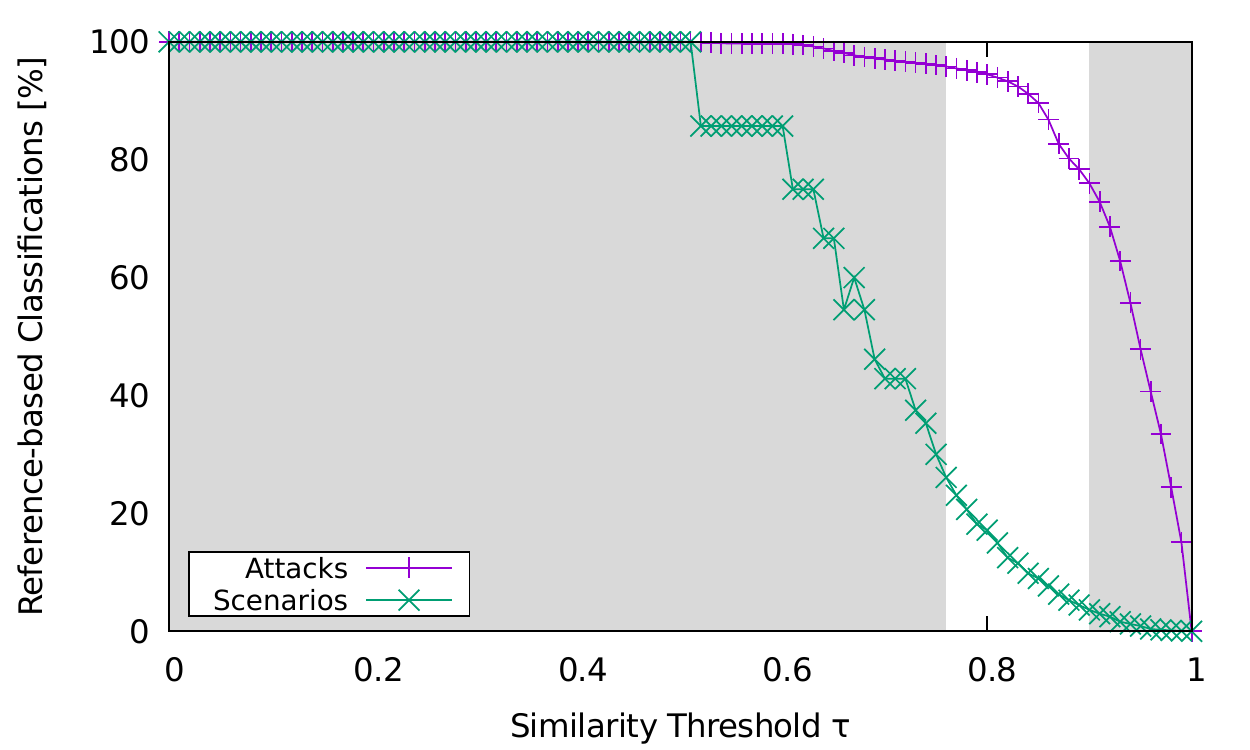}
		\caption{Portion of attacks classified by reference scenarios and portion\\of identified reference scenarios among all identified scenarios.}
		\label{fig:signature_attacks}
	\end{subfigure}
	\begin{subfigure}[b]{0.49\textwidth}
		\centering
		\includegraphics[width=\textwidth]{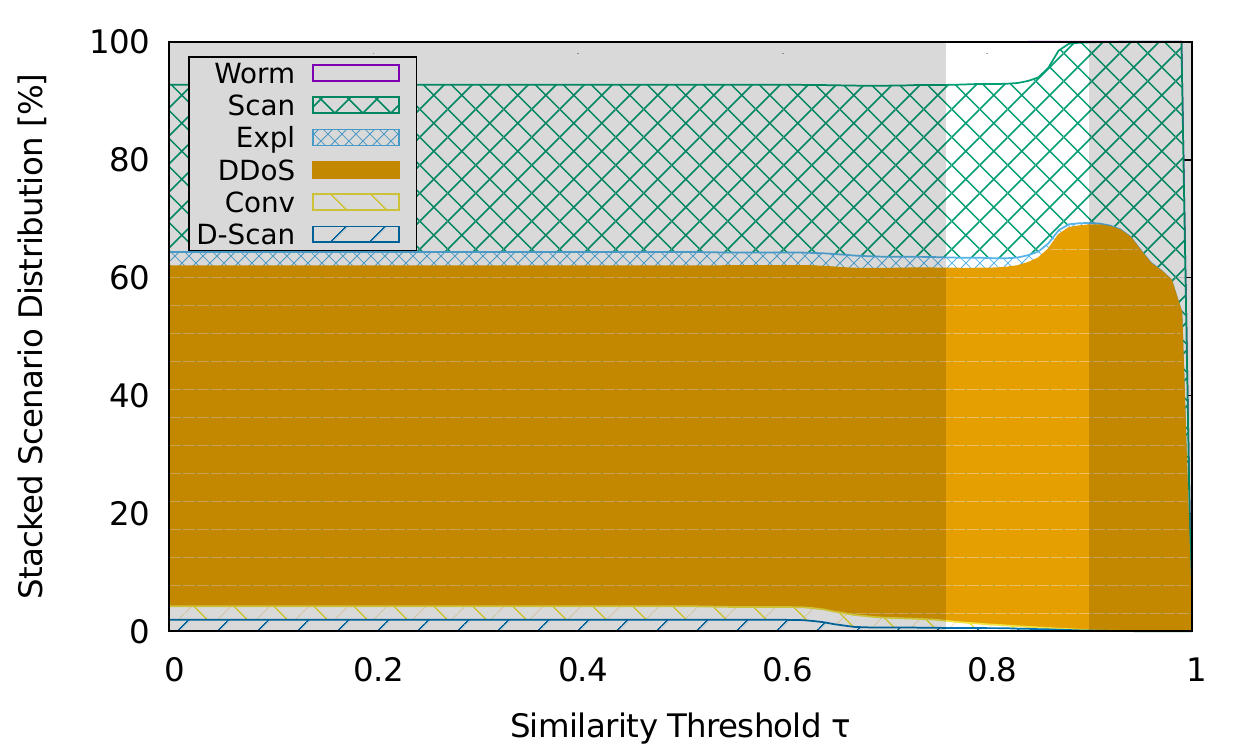}
		\caption{Portion of attacks per reference scenarios among all attacks classified by reference scenarios.}
		\label{fig:stacked_signatures}
	\end{subfigure}
	\caption{Classification of attacks using reference scenarios depending on similarity threshold $\tau$.}
	\label{fig:signature_numbers}
\end{figure*}

\subsection{Real-World Results}
\label{subsec:real_world_evaluation}

% Intro
After analyzing our scenario classification on artificial data for which we have the ground truth, we also apply it on the DShield real-world data (cf. Section~\ref{subsubsec:dshield_data}) and report the performance of our motif-based correlation algorithm for the detection of attack scenarios.
As reference scenarios, we use one attack with 100 host for each of the six attack scenarios in Section~\ref{subsubsec:attack_scenarios}. As the results of Section~\ref{subsec:signature_evaluation} indicate that the similarity threshold should be $\tau \in [0.76; 0.9]$, we mark the respective range for $\tau$ on the x-axis in all following figures. Choosing $\tau$ from this range is a prerequisite to distinguish attacks from the six reference scenarios.

For this real-world evaluation, we look both at classifying attacks with the help of reference scenarios and detecting scenarios with unsupervised clustering to compare them.

\subsubsection{Efficiency of Motif Signatures}
\label{subsubsec:real_eval_size}

An efficient data structure for the abstraction of attacks is necessary when sharing attack information and process them in a distributed manner~\cite{Locasto2005a,Yegneswaran2004}. To evaluate the compression rate of our motif signatures during the real-world experiments, we measured the total size required at different stages, i.e., in representation structures, for the 34,204 attacks in the DShield data set:

\begin{tabular}{ll}
	Alerts with all attributes: & 449 MB \\
	Alerts with IP/Port only: & 352 MB \\
	Motif signatures of attacks: & 5.1 MB 
\end{tabular}

Hence, the size was reduced to 1.12\% of the full alert data and 1.43\% of the alerts with relevant attributes only.

\subsubsection{Signature-based Classification}
\label{subsubsec:real_eval_signature}

% Experiment Setup
We first utilize reference-based clustering to identify attacks that we found in the DShield data set that can be classified using reference scenarios. If identified, we assign the attack to one of the six reference scenarios based on the similarity threshold $\tau$. For the attacks that could not be classified, we apply unsupervised clustering, which results in additional classes, i.e., unknown attack scenarios.

% Results
Figure~\ref{fig:signature_numbers} shows the performance of the reference-based classification depending on $\tau$. Figure~\ref{fig:signature_attacks} in particular illustrates how many attacks or scenarios have been classified or detected with the help of reference scenarios.
% - Number of classified attacks
For that, the curve labeled \textit{attacks} plots the portion of the 34,204 attacks that have been assigned to one of the six reference scenarios.
For a similarity threshold $\tau \leq 0.5$, all attacks are assigned a reference scenario. For larger $\tau$, the attacks have to match the reference scenarios more closely. It is likely that the attacks obtained from the DShield data set come with false-positive alerts. Therefore, these attacks cannot be assigned a reference scenario when close matches with the reference scenarios are required. Furthermore, the data set can contain attacks that are not covered by our six reference scenarios and will therefore not match any of them. In the marked range of $\tau$, however, we were able to classify at least 76\% and up to 96\% of the attacks in the DShield data set.
% Number of reference-based scenarios
In Figure~\ref{fig:signature_attacks} we investigate the relation between the number of scenarios detected through reference-based clustering and the number of scenarios detected through unsupervised clustering of the remaining attacks. We also plot the portion of the total detected scenarios which are identified with the help of a reference scenario. In the marked range of $\tau$, they are between 4\% and 26\%.

% Sizes of reference-based clusters
Figure~\ref{fig:stacked_signatures} shows the distribution of reference scenarios among the identified attacks. The portions of the scenarios are stacked, so the aggregation of all six scenarios is 100\%. As expected from a real-world data set, the most predominant attack scenarios are DDoS with in between 60\% and 69\% as well as Scan with in between 29\% and 31\%.
Although for each of the six reference scenarios there is at least one attack identified in the marked range of $\tau$, we have to note that not all reference scenarios can technically show up in the DShield data set. This is because the destination IP addresses are hashed and therefore it is not possible to observe attacks in which an individual host is both, an attacker and a victim. This excludes the scenarios worm, expl, and conv.
Considering this technical restriction, we conclude that in practice a large similarity threshold $\tau \leq 0.9$ should be chosen to avoid an unacceptable amount of false positive detections.

\subsubsection{Unsupervised Clustering}
\label{subsubsec:real_eval_dynamic}

\begin{figure*}[t]
	\centering
	\begin{subfigure}[b]{0.49\textwidth}
		\centering
		\includegraphics[width=\textwidth]{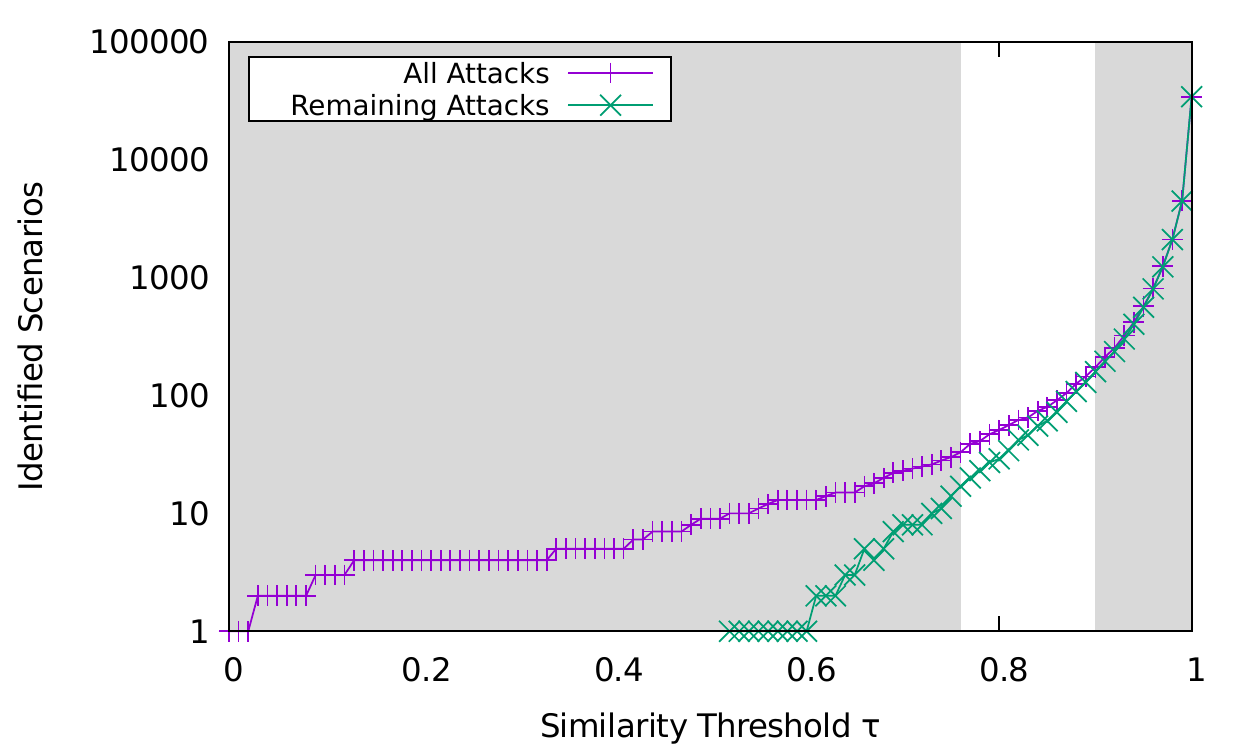}
		\caption{Detecting scenarios with unsupervised classification among all\\attacks and among remaining attacks from reference-based\\ classification.}
		\label{fig:signature_clusters}
	\end{subfigure}
	\begin{subfigure}[b]{0.49\textwidth}
		\centering
		\includegraphics[width=\textwidth]{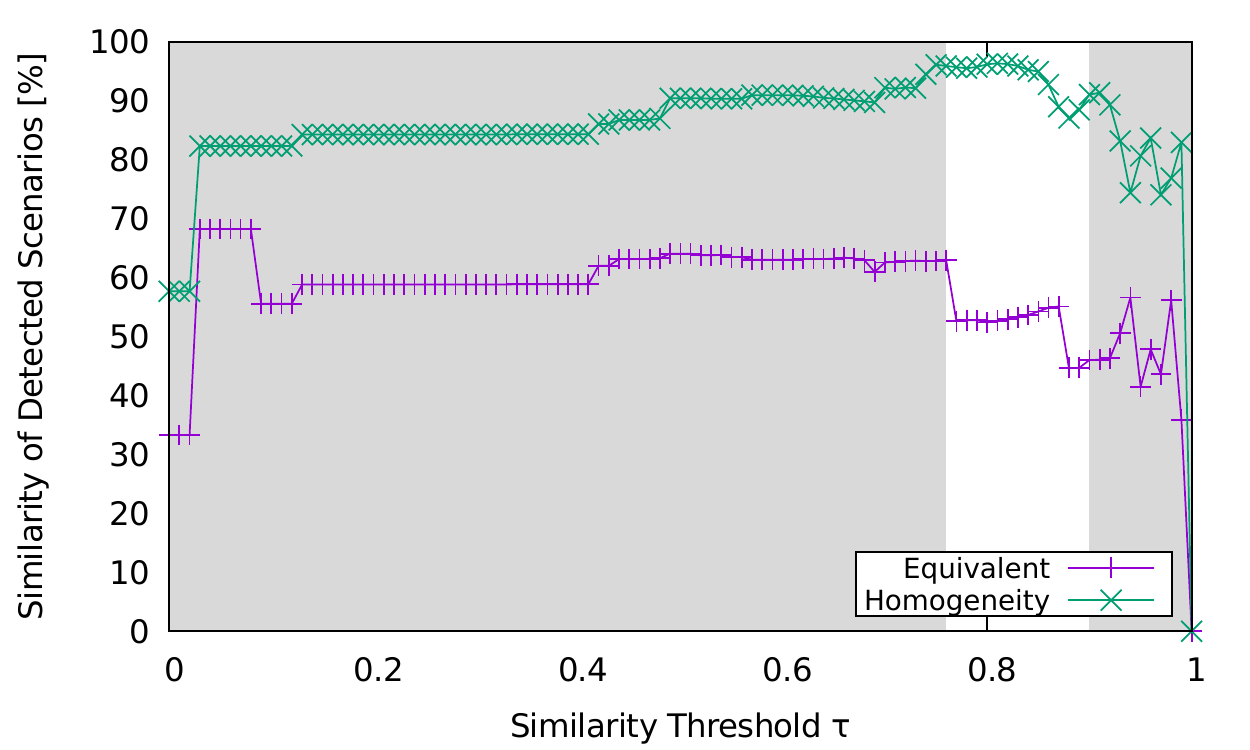}
		\caption{Similarity between scenario classes detected by reference scenarios and detected by unsupervised clustering.\\}
		\label{fig:dynamic_overlap}
	\end{subfigure}
	\caption{Identifying scenarios with reference-based and unsupervised classification depending on similarity threshold $\tau$.}
	\label{fig:dynamic_numbers}
\end{figure*}

% Experiment Setup
We now evaluate the unsupervised clustering of the complete DShield real-world data set. For that, we apply hierarchical clustering (cf. Section~\ref{subsubsec:dynamic_classfication}) on the DShield attacks and report the analysis of resulting classes in Figure~\ref{fig:dynamic_numbers} depending on the similarity threshold $\tau$.

% Number of scenarios
Figure~\ref{fig:signature_clusters} shows the number of detected scenarios for unsupervised clustering on a log-scale. Within the marked range of $\tau$, we see between 33 and 174 clusters. Defining scenarios by higher similarities, i.e., $\tau$, larger than 0.9, rapidly increases the number of scenarios and should only be used for fine-grained scenarios.
We also count the number of detected scenarios resulting from an unsupervised clustering of attacks that could not be assigned a reference scenario. We call these detected scenarios \textit{remaining} in Figure~\ref{fig:signature_clusters}.
The number of detected scenarios indicate that only searching for our six reference scenarios is not enough. Instead, it is important to also apply unsupervised clustering at least on attacks for which no reference scenario can be assigned.

% Overlap Similarity
We also compare the results of reference-based and unsupervised clustering. Our motivation is to find out for which value of the clustering parameter $\tau$ we get consistent results. As we have no ground-truth, we define two metrics to measure the similarity between the reference clusters $C^r_i \in \mathcal{C}^r$ for respective reference signatures $R^r_i \in \mathcal{R}^r$ and the unsupervised clusters $C^u_i \in \mathcal{C}^u$.

The metric \textbf{Equivalent}
	indicates how close candidates among the unsupervised clusters $C^u_i \in \mathcal{C}^u$ match our reference clusters $C^r_i \in \mathcal{C}^r$. For that, we find the best candidate cluster $C^u_i$ for every reference cluster $C^r_i$ based on the Jaccard index, i.e., intersection over union, which is calculated by $\frac{|C^u_i \cap C^r_i|}{|C^u_i \cup C^r_i|}$.
	An unsupervised cluster $C^u_i$ can be matched to at most one reference cluster $C^r_i$. The metric represents the average Jaccard metric for the best matches among all reference clusters $\mathcal{C}^r$.
	
The metric \textbf{Homogeneity}
	indicates the average accuracy for clusters $C^u_i \in \mathcal{C}^u$ that include at least one attack identified by the reference-based classification. In each of these clusters $C^u_i$, among all contained attacks, we measure the fraction of attacks from the scenario that is present in the cluster most frequently.
	The metric represents the average homogeneity among all these clusters weighted by their sizes.

The comparison between signature-based clustering and unsupervised clustering according to the metrics \textit{Equivalent} and \textit{Homogeneity} is shown in Figure~\ref{fig:dynamic_overlap}, depending on the similarity threshold $\tau$.
For the marked range of $\tau$, the \textit{Homogeneity} is between 87\% and 97\% and the \textit{Equivalent} is between 56\% and 72\%.
According to the experiment results of Section~\ref{subsubsec:real_eval_signature}, a similarity threshold $\tau$ close to 0.9 seems to be reasonable. Although the highest \textit{Homogeneity} is achieved for $\tau = 0.81$, which is in the lower half of the possible values for $\tau$, the \textit{Homogeneity} at $\tau = 0.9$ is still at 91\%. However, we note that between $0.81 \leq \tau \leq 0.9$ the metric \textit{Homogeneity} drops to 87\%. This drop correlates with the changes of the proportions among the different attack scenarios in Figure~\ref{fig:stacked_signatures}. From that perspective, $\tau= 0.9$ would be recommended to find attacks for our six scenario classes.

The two metrics \textit{Homogeneity} and \textit{Equivalent} for comparing signature-based clustering and unsupervised clustering indicate the following. Clustering attacks from unknown scenarios is done with high uncertainty. While unsupervised clustering is able to differentiate between attacks from different scenarios, it will not be able to perfectly classify a high variability of attacks without any previous knowledge. Hence, it is a good approach to provide as much reference scenarios as possible and to use unsupervised clustering to learn new attack scenarios and to create reference scenarios in a semi-supervised fashion.

\section{Conclusion}
\label{sec:conclusion}

% Approach
With our motif-based approach, we are able to calculate characteristic signatures of attacks. To achieve that, we process large amounts of attacks from alerts reported by Intrusion Detection Systems. Clustered alerts are converted into motif signatures that are an abstraction over the original attack. These signatures are of small sizes and thus can be compared very fast. With the help of this abstraction, we can identify known attack scenarios, detect similar attacks, and can even learn about new attack scenarios.
% Applications for motif-based attack signatures
Our attack correlation is not limited to central grouping of similar attacks or to label them with their respective attack scenario. It is also suitable for collaborative intrusion detection, where individual sensors could exchange only the small motif signatures instead of large amounts of alerts.

% Results
Our experiments indicate that the motif-based abstraction is a suitable representation of attacks to preserve the scenario characteristics, especially for attacks of different sizes. Using six representative attack scenarios, up to 96\% of attacks recorded in a real-world alert set were successfully classified. Based on both artificial and real-world experiments, we found the best performance could be achieved when clustering attacks with a similarity of at least 90\%. Furthermore, we highlighted how we can learn and detect new attack scenarios when operating our approach adaptively.

% Future work
For future work we plan to investigate more fine-grained attack scenarios. Also, we plan to use motif-based signatures for the detection of specific attack scenarios in large unclustered alert sets.

%
% The next two lines define the bibliography style to be used, and the bibliography file.
\balance
\bibliographystyle{./IEEEtran}
\bibliography{./motifs}

\end{document}